\documentclass[pra,aps,showpacs,twocolumn]{revtex4}
\usepackage{graphicx}
\usepackage{latexsym}
\usepackage{amsmath}
\usepackage[dvipdfm]{hyperref}

\input{tcilatex}
\begin{document}

\title{Multipartite fully entangled fraction}
\author{Jianwei Xu}
\email{xxujianwei@nwafu.edu.cn}
\affiliation{College of Science, Northwest A\&F University, Yangling, Shaanxi 712100,
China}
\date{\today}

\begin{abstract}
Abstract: Fully entangled fraction is a definition for bipartite states,
which is tightly related to bipartite maximally entangled states, and has
clear experimental and theoretical significance. In this work, we generalize
it to multipartite case, we call the generalized version multipartite fully
entangled fraction (MFEF). MFEF measures the closeness of a state to GHZ states. The analytical expressions of MFEF are very difficult to obtain except for very special states, however, we show that, the MFEF of
any state is determined by a system of finite-order polynomial equations.
Therefore, the MFEF can be efficiently numerically computed.
\end{abstract}

\maketitle

\bigskip

\section{Introduction}

Quantum entanglement is a crucial ingredient in quantum information
processing. In practice, the maximally entangled states are often the ideal
resource in many quantum information processing schemes \cite{1,Horodecki2009}. Fully
entangled fraction is a definition for bipartite states, which is tightly
related to bipartite maximally entangled states, and has clear experimental
and theoretical significance \cite{6,7,8,9,10}. In Ref.\cite{11}, the
analytical expressions are obtained for the fully entangled fraction of any
two-qubit states. In Ref.\cite{12} an upper bound of the fully entangled
fraction is obtained. In Ref.\cite{13} some analytical results have been
derived for some special states. In Ref.\cite{14}, the monogamy relations
for multiqubit states via the fully entangled fraction have been
investigated. Ref.\cite{4} shows that in $d\otimes
d^{\prime }$ ($2d\leq d^{\prime }$) system, there exist mixed maximally
entangled states. Based on this fact of Ref.\cite{4}, Ref.\cite{Zhao2015} studies the maximally
entangled states and fully entangled fraction in general $d\otimes d^{\prime
}$ system.

In this work, we generalize the definition of fully entangled fraction to
multipartite case, we call the generalized version multipartite fully
entangled fraction (MFEF). Our definition of MFEF is tightly related to the GHZ states. GHZ states are a class of important multipartite entangled states which play vital roles in many experiments either testing the quantum formalism or realizing quantum information processing \cite{Horodecki2009}. The bipartite maximally entangled pure states can be viewed as the bipartite case of GHZ states. Similar to the case of fully entangled fraction, we show
that the analytical expressions of MFEF are very difficult to obtain except
for very special states. However, we prove that, the MFEF of any state is
determined by a system of finite-order polynomial equations. Therefore, the
MFEF can be efficiently numerically computed.

This paper is organized as follows. In section 2, we give the definition of
MFEF, a lower bound and an upper bound of MFEF, and give the expressions of
MFEF for a class of special pure states. In section 3, we study the MFEF for
$N$-qubit states. In section 4, we show that the MFEF of any state is
determined by a system of finite-order polynomial equations, therefore, the
MFEF can be efficiently numerically computed. In section 5 we give a summary.

\section{The definition of MFEF}

Consider the $N$-partite $(N\geq 2)$ system $A_{1}A_{2}...A_{N}$, its
subsystems $\{A_{l}\}_{l=1}^{N}$ are all $d$-dimensional and correspond to
the Hilbert space $H$. We define the multipartite fully entangled fraction
(MFEF) of the $N$-partite state $\rho $ on $H^{\otimes N}$ (we also write $%
H^{\otimes N}=d^{\otimes N}$ for emphasizing $dimH=d$) as

\begin{eqnarray}
F(\rho )=\max_{U_{1},...,U_{N}}\langle \phi |(\otimes
_{l=1}^{N}U_{l}^{+})\rho (\otimes _{l=1}^{N}U_{l})|\phi \rangle ,
\end{eqnarray}

where max runs over all local $d\times d$ unitary matrices $%
U_{1},U_{2},...,U_{N},$ $+$ denotes adjoint, and $|\phi \rangle $ is the GHZ
state for given orthonormal basis $\{|i\rangle \}_{i=1}^{d}$ of $H$,
\begin{eqnarray}
|\phi \rangle =\frac{1}{\sqrt{d}}\sum_{i=1}^{d}|ii...i\rangle .
\end{eqnarray}
Obviously, the bipartite maximally entangled pure states can be viewed as the bipartite case of GHZ states, and the definition of fully entangled fraction can be viewed as the bipartite case of MFEF.
The intuitive meaning of MFEF is that it measures the closeness of a state with respect to the GHZ states.

\textbf{Theorem 1.}The MFEF of the $N$-partite state $\rho $ on $d^{\otimes
N}$ satisfies that
\begin{eqnarray}
\frac{1}{d^{N}}\leq F(\rho )\leq p_{\max }(\rho )\leq \sqrt{tr(\rho ^{2})}%
\leq 1; \\
F(\rho )=1\Leftrightarrow \rho \text{\ is a GHZ state}; \\
F(\rho )=\frac{1}{d^{N}}\Leftrightarrow \rho =\frac{I^{\otimes N}}{d^{N}}.
\end{eqnarray}
In Eq.(3), $p_{\max }(\rho )$ is the maximal eigenvalues of $%
\rho $, in Eq.(5), $I$ is the identity operator.

\textbf{Proof.} We prove this theorem in the similar way of the proof for
Theorem 2 in Ref.\cite{13}. Let $\rho =\sum_{i=1}^{d^{N}}p_{i}|\psi _{i}\rangle
\langle \psi _{i}|$ be the eigendecomposition,
\begin{eqnarray}
\langle \phi |(\otimes _{i=1}^{N}U_{i}^{+})\rho (\otimes
_{i=1}^{N}U_{i})|\phi \rangle =\sum_{i}p_{i}q_{i}  \notag \\
\leq p_{\max }\leq \sqrt{\sum_{i}p_{i}^{2}}=\sqrt{tr(\rho ^{2})}\leq 1,
\end{eqnarray}
with
\begin{eqnarray}
q_{i}=\langle \phi |(\otimes _{i=1}^{N}U_{i}^{+})|\psi _{i}\rangle \langle
\psi _{i}|(\otimes _{i=1}^{N}U_{i})|\phi \rangle .
\end{eqnarray}
If $\sum_{i}p_{i}q_{i}=1$, then $p_{\max }=1,q_{\max }=1$, thus $\rho =|\psi
\rangle \langle \psi |$ is pure, and
\begin{eqnarray}
\langle \phi |(\otimes _{i=1}^{N}U_{i}^{+})|\psi \rangle \langle \psi
|(\otimes _{i=1}^{N}U_{i})|\phi \rangle =1,
\end{eqnarray}
it follows that $\rho =(\otimes _{i=1}^{N}U_{i})|\phi \rangle \langle \phi
|(\otimes _{i=1}^{N}U_{i}^{+})$ for any unitary matrices $%
U_{1},U_{2},...,U_{N}.$

Using the method of Lagrange multipliers, it can be shown that the minimum
of $\sum_{i}p_{i}q_{i}$ is $\frac{1}{d^{N}}$ by $p_{i}=q_{i}=\frac{1}{d^{N}}$
for all $i$, hence Eq.(5) holds. We then end this proof.

\textbf{Theorem 2.} The MFEF of the pure state
\begin{eqnarray}
|\psi \rangle =\sum_{i=1}^{d}\sqrt{p_{i}}|ii...i\rangle
\end{eqnarray}
is
\begin{eqnarray}
F(\psi )=(\sum_{i=1}^{d}\sqrt{p_{i}})^{2},
\end{eqnarray}
where, $\{|i\rangle \}_{i=1}^{d}$ is an orthonormal basis of $H$, $p_{i}\geq
0,$ $\sum_{i=1}^{d}p_{i}=1.$

\textbf{Proof.} First note that
\begin{eqnarray}
\sum_{j=1}^{d}|(U_{1})_{ji}(U_{2})_{ji}...(U_{N})_{ji}| \ \ \ \ \ \ \ \ \ \
\ \ \ \ \ \ \ \ \ \ \ \ \ \ \ \   \notag \\
\leq \sqrt{\sum_{j=1}^{d}|(U_{1})_{ji}|^{2}%
\sum_{j=1}^{d}|(U_{2})_{ji}|^{2}...\sum_{j=1}^{d}|(U_{N})_{ji}|^{2}}\leq 1.
\end{eqnarray}
Let $\{|i\rangle \}_{i=1}^{d}=\{|j\rangle \}_{j=1}^{d}=\{|j_{1}\rangle
\}_{j_{1}=1}^{d}$ denote the same orthonormal basis of $H$, then
\begin{eqnarray}
F(\psi )=\max |\langle \psi |(\otimes _{l=1}^{N}U_{l})|\phi \rangle |^{2} \
\ \ \ \ \ \ \ \ \ \ \ \ \ \ \ \ \ \ \ \ \ \ \ \ \ \ \   \notag \\
=\frac{1}{d}\max |\sum_{ij=1}^{d}\sqrt{p_{i}}\langle ii...i|\cdot
U_{1}|j\rangle U_{2}|j\rangle ...U_{N}|j\rangle |^{2} \ \ \ \ \ \ \ \ \ \ \
\notag \\
=\frac{1}{d}\max |\sum_{ij,j_{1}...j_{N}}\sqrt{p_{i}}\langle
ii...i|(U_{1})_{jj_{1}}|j_{1}\rangle...(U_{N})_{jj_{N}}|j_{N}\rangle |^{2}
\notag \\
=\frac{1}{d}\max |\sum_{i=1}^{d}\sqrt{p_{i}}%
[\sum_{j=1}^{d}(U_{1})_{ji}(U_{2})_{ji}...(U_{N})_{ji}]|^{2} \ \ \ \ \ \ \ \
\ \ \ \   \notag \\
\leq \frac{1}{d}\max [\sum_{i=1}^{d}\sqrt{p_{i}}%
|\sum_{j=1}^{d}(U_{1})_{ji}(U_{2})_{ji}...(U_{N})_{ji}|]^{2} \ \ \ \ \ \ \ \
\ \ \ \   \notag \\
\leq \frac{1}{d}(\sum_{i=1}^{d}\sqrt{p_{i}})^{2}. \ \ \ \ \ \ \ \ \ \ \ \ \
\ \ \ \ \ \ \ \ \ \ \ \ \ \ \ \ \ \ \ \ \ \ \ \ \ \ \ \ \ \ \ \ \ \ \ \
\end{eqnarray}

Let $U_{1}=U_{2}=...=U_{N}=I$, we have $|\langle \psi |\phi \rangle |^{2}=%
\frac{1}{d}(\sum_{i=1}^{d}\sqrt{p_{i}})^{2}.$ We then end this proof.

\section{MFEF of $N$-qubit states}

In this section, we give a special $N$-qubit states which allow analytical
MFEF, and investigate the MFEF for arbitrary $N$-qubit states.

\textbf{Theorem 3.} The MFEF of the two-qubit state
\begin{eqnarray}
\rho =\frac{1}{2^{\otimes N}}(I^{\otimes N}+c\sigma _{3}^{\otimes N})
\end{eqnarray}
is
\begin{eqnarray}
F(\rho )=\frac{1+|c|}{2^{N}},
\end{eqnarray}
where, $\sigma _{3}=\left(
\begin{array}{cc}
1 & 0 \\
0 & -1%
\end{array}%
\right) $, $c$ is a real number satisfying $|c|\leq 1.$

\textbf{Proof.} For the Pauli matrices
\begin{eqnarray}
\sigma _{1}=\left(
\begin{array}{cc}
0 & 1 \\
1 & 0%
\end{array}%
\right) ,\sigma _{2}=\left(
\begin{array}{cc}
0 & -i \\
i & 0%
\end{array}%
\right) ,\sigma _{3}=\left(
\begin{array}{cc}
1 & 0 \\
0 & -1%
\end{array}%
\right) ,
\end{eqnarray}
and any $2\times 2$ unitary matrix $U$, there exists a $3\times 3$ real
orthogonal matrix $O$ such that \cite{Horn1985}
\begin{eqnarray}
U^{+}(\sum_{j=1}^{3}r_{j}\sigma _{j})U=\sum_{jk=1}^{3}O_{jk}r_{k}\sigma _{j},
\end{eqnarray}
for any real numbers $\{r_{j}\}_{j=1}^{3}$.

For given $2\times 2$ unitary matrices $\{U_{l}\}_{l=1}^{N},$we denote the
corresponding real orthogonal matrices as $\{O^{(l)}\}_{l=1}^{N}.$

For the state in Eq.(13),
\begin{eqnarray}
F(\rho )=\frac{1}{2^{N}}\{1+\max [c\langle \phi |\otimes
_{l=1}^{N}(U_{l}^{+}\sigma _{3}U_{l})|\phi \rangle ]\}.
\end{eqnarray}
Taking Eq.(16) into above equation, with direct computations, we get
\begin{eqnarray}
2\langle \phi |\otimes _{i=1}^{N}(U_{i}^{+}\sigma _{3}U_{i})|\phi \rangle \
\ \ \ \ \ \ \ \ \ \ \ \ \ \ \ \ \ \ \ \ \ \ \ \ \ \ \ \ \ \ \ \ \ \ \ \ \
\notag \\
=O_{33}^{(1)}O_{33}^{(2)}...O_{33}^{(N)}+(-1)^{N}O_{33}^{(1)}O_{33}^{(2)}...O_{33}^{(N)} \ \ \ \ \ \ \ \ \ \ \ \ \ \ \
\notag \\
+(O_{13}^{(1)}+iO_{23}^{(1)})...(O_{13}^{(N)}+iO_{23}^{(N)}) \ \ \ \ \ \ \ \
\ \ \ \ \ \ \ \ \ \ \ \ \ \ \   \notag \\
+(O_{13}^{(1)}-iO_{23}^{(1)})...(O_{13}^{(N)}-iO_{23}^{(N)}). \ \ \ \ \ \ \
\ \ \ \ \ \ \ \ \ \ \ \ \ \ \
\end{eqnarray}
Consequently,
\begin{eqnarray}
|c\langle \phi |\otimes _{i=1}^{N}(U_{i}^{+}\sigma _{3}U_{i})|\phi \rangle |
\ \ \ \ \ \ \ \ \ \ \ \ \ \ \ \ \ \ \ \ \ \ \ \ \ \ \ \ \ \ \ \ \ \ \ \ \
\notag \\
\leq |c|\lbrack
|O_{33}^{(1)}...O_{33}^{(N)}|+|(O_{13}^{(1)}+iO_{23}^{(1)})...(O_{13}^{(N)}+iO_{23}^{(N)})|]
\notag \\
\leq |c|\cdot \sqrt{\sum_{j}|O_{j3}^{(1)}|^{2}}...\sqrt{%
\sum_{j}|O_{j3}^{(N)}|^{2}}=|c|. \ \ \ \ \ \ \ \ \ \ \ \ \ \ \ \ \ \
\end{eqnarray}
Conversely, let $O_{13}^{(1)}=O_{13}^{(2)}=...=O_{13}^{(N-1)}=1$, and $%
O_{13}^{(N)}=sign(c) $, all above equalities hold. We then end this proof.

We next investigate the MFEF for arbitrary $N$-qubit states. The $2\times 2$
unitary matrices $\{U^{(l)}\}_{l=1}^{N}$ can be expressed as
\begin{eqnarray}
U^{(l)}=x_{0}^{(l)}I+i\sum_{j=1}^{3}x_{j}^{(l)}\sigma _{j}=\sum_{\mu
=0}^{3}x_{\mu }^{(l)}i^{g(\mu )}\sigma _{\mu },
\end{eqnarray}
where $\{x_{\mu }^{(l)}\}_{\mu =0}^{3}$ are real numbers,
\begin{eqnarray}
\sum_{\mu =0}^{3}(x_{\mu }^{(l)})^{2}=1, for \ all \ l, \ \ \ \ \ \ \ \ \ \
\ \ \ \ \ \ \ \  \\
\sigma _{0}=I, \ \ \ \ \ \ \ \ \ \ \ \ \ \ \ \ \ \ \ \ \ \ \ \ \ \ \ \ \ \ \
\ \ \ \ \  \\
g(\mu )=0 \ when \ \mu =0, \ g(\mu )=1 \ when \ \mu =1,2,3.
\end{eqnarray}
As a result,
\begin{eqnarray}
(\otimes _{l=1}^{N}U_{l})|\phi \rangle =\sum_{\mu _{1}\mu _{2}...\mu
_{N}=0}^{3}x_{\mu _{1}}^{(1)}x_{\mu _{2}}^{(2)}...x_{\mu _{N}}^{(N)} \ \ \ \
\ \ \ \ \ \ \ \ \ \ \ \   \notag \\
\cdot i^{g(\mu_{1})+g(\mu _{2})+...+g(\mu _{N})}\sigma _{\mu _{1}}\otimes
\sigma _{\mu _{2}}...\sigma _{\mu _{N}}|\phi \rangle \  \\
=\sum_{\mu _{1}\mu _{2}...\mu _{N}=0}^{3}x_{\mu _{1}}^{(1)}x_{\mu
_{2}}^{(2)}...x_{\mu _{N}}^{(N)}|\phi _{\mu _{1}\mu _{2}...\mu _{N}}\rangle,
\ \ \ \
\end{eqnarray}
where
\begin{eqnarray}
|\phi _{\mu _{1}...\mu _{N}}\rangle =i^{g(\mu _{1})+...+g(\mu _{N})}\sigma
_{\mu _{1}}\otimes ...\otimes\sigma _{\mu _{N}}|\phi \rangle .
\end{eqnarray}
Consequently,
\begin{eqnarray}
\langle \phi |(\otimes _{i=1}^{N}U_{i}^{+})\rho (\otimes
_{i=1}^{N}U_{i})|\phi \rangle \ \ \ \ \ \ \ \ \ \ \ \ \ \ \ \ \ \ \ \ \ \ \
\   \notag \\
=\sum_{\nu _{l};\mu _{l}}x_{\nu _{1}}^{(1)}...x_{\nu_{N}}^{(N)}x_{\mu
_{1}}^{(1)}...x_{\mu _{N}}^{(N)}\langle \phi _{\nu _{1}...\nu _{N}}|\rho
|\phi _{\mu _{1}...\mu _{N}}\rangle  \notag \\
=\sum_{\nu _{l};\mu _{l}}x_{\nu _{1}}^{(1)}...x_{\nu_{N}}^{(N)}x_{\mu
_{1}}^{(1)}...x_{\mu _{N}}^{(N)}R _{\nu _{1}...\nu _{N};\mu _{1}...\mu_{N}}
, \ \ \ \ \
\end{eqnarray}
where, $R_{\nu _{1}...\nu _{N},\mu _{1}...\mu _{N}}$ is the real part of $%
\langle \phi _{\nu _{1}...\nu _{N}}|\rho |\phi _{\mu _{1}...\mu _{N}}\rangle
,$
\begin{eqnarray}
R_{\nu _{1}...\nu _{N},\mu _{1}...\mu _{N}} \ \ \ \ \ \ \ \ \ \ \ \ \ \ \ \
\ \ \ \ \ \ \ \ \ \ \ \ \ \ \ \ \ \ \ \ \ \ \ \ \ \   \notag \\
=\frac{1}{2}[\langle \phi _{\nu _{1}...\nu _{N}}|\rho |\phi _{\mu _{1}...\mu
_{N}}\rangle +\langle \phi _{\mu _{1}...\mu _{N}}|\rho|\phi _{\nu _{1}...\nu
_{N}}\rangle ], \\
R_{\nu _{1}...\nu _{N},\mu _{1}...\mu _{N}}=R_{\mu _{1}...\mu _{N},\nu
_{1}...\nu _{N}}. \ \ \ \ \ \ \ \ \ \ \ \ \ \ \ \ \ \ \
\end{eqnarray}

Let $\{\lambda _{l}\}_{l=1}^{N}$ be the Lagrange multipliers, and let
\begin{eqnarray}
L=\sum_{\nu _{l};\mu _{l}}x_{\nu _{1}}^{(1)}...x_{\nu _{N}}^{(N)}x_{\mu
_{1}}^{(1)}...x_{\mu _{N}}^{(N)}R_{\nu _{1}...\nu _{N},\mu _{1}...\mu _{N}}
\notag \\
-\sum_{l=1}^{N}\lambda _{l}[\sum_{\mu =0}^{3}(x_{\mu }^{(l)})^{2}-1],
\end{eqnarray}
then the extremum conditions yield that
\begin{eqnarray}
\frac{1}{2}\frac{\partial L}{\partial x_{\nu _{l}}^{(l)}}=\sum_{no \ \nu
_{l}}x_{\nu_{1}}^{(1)}...x_{\nu
_{l-1}}^{(l-1)}x_{\nu_{l+1}}^{(l+1)}...x_{\nu _{N}}^{(N)}x_{\mu
_{1}}^{(1)}...x_{\mu _{N}}^{(N)}  \notag \\
\cdot R_{\nu _{1}...\nu _{N},\mu _{1}...\mu _{N}}-\lambda _{l}x_{\nu
_{l}}^{(l)}=0, \ \ \ \ \ \ \ \ \ \
\end{eqnarray}
for all $l$ and all $\nu_{l}=0,1,2,3.$ Taking Eq.(31) into Eq.(27), we get
\begin{eqnarray}
\langle \phi |(\otimes _{i=1}^{N}U_{i}^{+})\rho (\otimes
_{i=1}^{N}U_{i})|\phi \rangle =\lambda _{l}.
\end{eqnarray}
Note that, in general, Eq.(31) leads to many (but finite) solutions of $%
\{x_{\mu _{l}}^{(l)},\lambda _{l}\}$, we should take the maximum of Eq.(32)
for all these $\lambda _{l}.$ We conclude this result as Theorem 4 below.

\textbf{Theorem 4.} For any $N$-qubit state $\rho $, the real tensor $R_{\nu
_{1}...\nu _{N},\mu _{1}...\mu _{N}}$ is defined in Eqs.(28,26), then the
MFEF of $\rho $ is
\begin{eqnarray}
F(\rho )=\max_{l=1}^{N}\{\max [\lambda _{l}:Eqs.(31),Eqs.(21)]\}.
\end{eqnarray}

We remark that, when $N=2$, Eq.(33) recovers the two-qubit case in Ref.\cite{11} (noticing that when $N=2$ we can always equivalently let $U _{1}=I$).

\section{MFEF of $N$-qudit states}

From the Theorem 4 for the \emph{N}-qubit ($d=2$) case, we want to know
whether the MFEF of any \emph{N}-qudit state can also be obtained by solving
a system of finite-order polynomial equations. In this section we show this
is true. To this aim, we use the generators of su($d$) Lie algebra to represent unitary
matrices on $H$. Consider the $d^{2}$ Hermitian operators \cite{15,16} on $%
H: $
\begin{eqnarray}
\{\sigma _{\mu }\}_{\mu =0}^{d^{2}-1}=\{\sigma _{0}=\sqrt{\frac{2}{d}}%
I,\sigma _{1},...,\sigma _{d^{2}-1}\}, \ \ \ \ \ \ \ \ \ \ \ \ \ \ \ \ \ \ \
\  \\
\{\sigma _{i}\}_{i=1}^{d^{2}-1}=\cup\{(|k\rangle \langle j|+|j\rangle
\langle k|)_{j\neq k}\}_{jk=1}^{d} \ \ \ \ \ \ \ \ \ \ \ \ \ \ \ \ \ \ \ \
\notag \\
\cup\{-i(|k\rangle \langle j|-|j\rangle \langle k|)_{l<k}\}_{jk=1}^{d} \ \ \
\ \ \ \ \ \ \ \ \ \ \ \ \ \ \ \ \   \notag \\
\cup\{\sqrt{\frac{2}{j(j+1)}}(\Sigma _{k=1}^{j}|k\rangle \langle
k|-j|j+1\rangle \langle j+1|)\}_{j=1}^{d-1}. \ \ \
\end{eqnarray}
It is shown that
\begin{eqnarray}
tr\sigma _{0}=\sqrt{2d}. \ \ \ \ Tr\sigma _{j}=0, \ j\in\{1,...,d^{2}-1\}, \\
tr(\sigma _{\mu }\sigma _{\nu })=2\delta _{\mu \nu }, \ \ \mu ,\nu \in
\{0,1,...,d^{2}-1\}.
\end{eqnarray}
For $i,j,k\in \{0,1,...,d^{2}-1\},$
\begin{eqnarray}
[\sigma _{i},\sigma _{j}]=2i\sum_{k=1}^{d^{2}-1}f_{ijk}\sigma _{k}, \\
\{\sigma _{i},\sigma _{j}\}=\frac{4}{d}\delta
_{ij}I+2\sum_{k=1}^{d^{2}-1}d_{ijk}\sigma _{k}, \\
\sigma _{i}\sigma _{j}=\frac{2}{d}\delta
_{ij}I+\sum_{k=1}^{d^{2}-1}(d_{ijk}+if_{ijk})\sigma _{k}, \\
f_{ijk}=\frac{1}{4i}Tr([\sigma _{i},\sigma _{j}]\sigma _{k}), \\
d_{ijk}=\frac{1}{4}Tr(\{\sigma _{i},\sigma _{j}\}\sigma _{k}),
\end{eqnarray}
where $[\sigma _{i},\sigma _{j}]$, $\{\sigma _{i},\sigma _{j}\}$ are
commutator and anticommutator, $f_{ijk},d_{ijk}$ are called structure
constants, $f_{ijk}$ are completely antisymmetric and $d_{ijk}$ completely
symmetric. When $d=2$, $\{\sigma _{i}\}_{i=1}^{3}$ are well known Pauli
operators, $d_{ijk}=0$ and $f_{ijk}$ the permutation symbol.

All $d\times d$ complex matrices forms a $d^{2}$-dimensional complex Hilbert
space $\mathcal{H}$ equipped the inner product $\langle M_{1}|M_{2}\rangle
=tr(M_{1}^{+}M_{2})$ for any $M_{1},M_{2}\in \mathcal{H}$. Note that Eq.(34)
is an orthonormal basis of $\mathcal{H}$, so any $d\times d$ complex matrix
can be expressed in this basis with complex coefficients. We express $%
d\times d$ Unitary matrix $U$ as
\begin{eqnarray}
U=\sum_{\mu =0}^{d^{2}-1}z_{\mu }\sigma _{\mu }=z_{0}\sigma
_{0}+\sum_{j=1}^{d^{2}-1}z_{j}\sigma _{j},z_{\mu }\in C,
\end{eqnarray}
where $\{z_{\mu }\}$ are all complex numbers satisfying the condition $%
UU^{+}=I$. Using Eq.(43), after some algebras, $UU^{+}=I$ leads to
\begin{eqnarray}
UU^{+}=(|z_{0}|^{2}+\sum_{j=1}^{d^{2}-1}|z_{j}|^{2})\frac{2}{d}I  \ \ \ \ \ \ \ \ \ \ \ \ \  \ \ \ \ \ \ \ \ \ \ \ \ \ \ \ \ \ \ \ \    \nonumber \\
+\sum_{k=1}^{d^{2}-1}[\sqrt{\frac{2}{d}}(z_{0}^{\ast
}z_{k}+z_{0}z_{k}^{\ast })+\sum_{ij=1}^{d^{2}-1}z_{i}z_{j}^{\ast
}(d_{ijk}+if_{ijk})]\sigma _{k}, \ \
\end{eqnarray}
thus
\begin{eqnarray}
|z_{0}|^{2}+\sum_{j=1}^{d^{2}-1}|z_{j}|^{2} =\frac{d}{2},  \\
\sqrt{\frac{2}{d}}(z_{0}^{\ast }z_{k}+z_{0}z_{k}^{\ast
})+\sum_{ij=1}^{d^{2}-1}z_{i}z_{j}^{\ast }(d_{ijk}+if_{ijk}) =0.
\end{eqnarray}%
Let $z_{\mu }=x_{\mu }+iy_{\mu }$ with $x_{\mu },y_{\mu }$ real, and notice
that $f_{ijk}$ are completely antisymmetric and $d_{ijk}$ completely
symmetric, then Eqs.(45,46) become
\begin{eqnarray}
\sum_{\mu =0}^{d^{2}-1}(x_{\mu }^{2}+y_{\mu }^{2})-\frac{d}{2} =0, \ \ \ \ \ \ \ \     \\
2\sqrt{\frac{2}{d}}(x_{0}x_{k}+y_{0}y_{k})+%
\sum_{ij=1}^{d^{2}-1}[(x_{i}x_{j}+y_{i}y_{j})d_{ijk} \nonumber \\
+2x_{i}y_{j}f_{ijk}] =0. \ \ \ \ \ \
\end{eqnarray}%
Let $z_{\mu _{l}}^{(l)}=x_{\mu _{l}}^{(l)}+iy_{\mu _{l}}^{(l)}$ be
the complex numbers corresponding to $U^{(l)}$ as in Eqs.(47,48), then
\begin{eqnarray}
\langle \phi |(\otimes _{l=1}^{N}U_{l}^{+})\rho (\otimes
_{l=1}^{N}U_{l})|\phi \rangle   \ \ \ \ \ \ \ \ \ \ \ \ \ \ \ \ \  \ \ \ \ \ \ \ \ \ \ \ \ \ \ \ \ \       \nonumber \\
=\sum_{\nu _{l},\mu _{l}}z_{\nu _{1}}^{(1)\ast }...z_{\nu _{N}}^{(N)\ast }z_{\mu _{1}}^{(1)}...z_{\mu
_{N}}^{(N)}\langle \phi _{\mu _{1}\mu _{2}...\mu _{N}}|\rho |\phi _{\mu
_{1}\mu _{2}...\mu _{N}}\rangle ,  \
\end{eqnarray}
where
\begin{eqnarray}
|\phi _{\mu _{1}\mu _{2}...\mu _{N}}\rangle =\sigma _{\mu _{1}}\otimes
...\sigma _{\mu _{N}}|\phi \rangle .
\end{eqnarray}%
Let
\begin{eqnarray}
L=\langle \phi |(\otimes _{l=1}^{N}U_{l}^{+})\rho (\otimes
_{l=1}^{N}U_{l})|\phi \rangle   \ \ \ \ \ \ \ \ \ \ \  \ \ \ \ \ \ \ \ \ \ \ \ \ \ \    \nonumber \\
-\sum_{l=1}^{N}\lambda _{l}\{\sum_{\mu
_{l}=0}^{d^{2}-1}[(x_{\mu _{l}}^{(l)})^{2}+(x_{\mu _{l}}^{(l)})^{2}]-\frac{d}{2}\}    \ \ \ \ \ \ \  \ \ \ \ \ \ \   \nonumber \\
-\sum_{l=1}^{N}\sum_{k=1}^{d^{2}-1}\tau _{lk}\{2\sqrt{\frac{2}{d}}%
(x_{0}^{(l)}x_{k}^{(l)}+y_{0}^{(l)}y_{k}^{(l)})   \ \ \ \ \ \ \  \ \ \ \    \nonumber \\
+\sum_{ij=1}^{d^{2}-1}[(x_{i}^{(l)}x_{j}^{(l)}+y_{i}^{(l)}y_{j}^{(l)})d_{ijk}+2x_{i}^{(l)}y_{j}^{(l)}f_{ijk}]\},
\end{eqnarray}%
then
\begin{eqnarray}
\frac{\partial L}{\partial x_{\mu _{l}}^{(l)}}=0, \ \
 \frac{\partial L}{\partial y_{\mu _{l}}^{(l)}}=0, \ \
 \frac{\partial L}{\partial \lambda _{l}}=0, \ \
 \frac{\partial L}{\partial \tau _{lk}}=0,
\end{eqnarray}%
constitute a system of finite-order polynomial equations in variables $%
\{x_{\mu _{l}}^{(l)},y_{\mu _{l}}^{(l)},\lambda _{l},\tau _{lk}\}$. Once we get
$\{x_{\mu _{l}}^{(l)},y_{\mu _{l}}^{(l)},\lambda _{l},\tau _{lk}\}$ from
Eqs.(52), taking them into $\langle \phi |(\otimes _{l=1}^{N}U_{l}^{+})\rho
(\otimes _{l=1}^{N}U_{l})|\phi \rangle $, we can get the MFEF. Note that, in
general, Eqs.(52) leads to many (but finite) solutions of $\{x_{\mu
_{l}}^{(l)},y_{\mu _{l}}^{(l)},\lambda _{l},\tau _{lk}\}$, we should take
the maximum of $\langle \phi |(\otimes _{l=1}^{N}U_{l}^{+})\rho (\otimes
_{l=1}^{N}U_{l})|\phi \rangle $ for all these $\{x_{\mu _{l}}^{(l)},y_{\mu
_{l}}^{(l)},\lambda _{l},\tau _{lk}\}.$

\section{Summary}

We generalized the definition of bipartite fully entangled fraction to the
multipartite case, we called the generalized version multipartite fully
entangled fraction (MFEF). MFEF is defined with respect to the multipartite
GHZ states then it measures the closeness of a state to GHZ states. We gave two classes of states which allow analytical MFEF, explored the bounds of MFEF. For $N$-qubit states, the optimization of MFEF
is relatively simple, we provided a calculation scheme. Although the
analytical MFEF are very hard to get for general states, we proved that, the
MFEF of any state can be efficiently numerically computed.

This work was supported by the Chinese Universities Scientific Fund (Grant
No.2014YB029) and the National Natural Science Foundation of China (Grant
No.11347213). The author thanks Kai-Liang Lin for helpful discussions.

\end{document}